\documentclass[10pt,twocolumn,letterpaper]{article}

\usepackage{iccv}              

\usepackage{soul}
\usepackage{float} 

\definecolor{iccvblue}{rgb}{0.21,0.49,0.74}
\usepackage[pagebackref,breaklinks,colorlinks,allcolors=iccvblue]{hyperref}
\usepackage{algorithm}
\usepackage{algpseudocode}
\def\paperID{15} 
\def\confName{ICCV}
\def\confYear{2025}

\title{DIFFRACT: Diffusion-based Restoration via Adaptive Control and Thresholding for Diffraction Imaging}

\author{Nikolay Falaleev\\
Fanis Technologies\\
London, UK\\
{\tt\small n.falaleev@fanis.ai}
\and
Nikolai Orlov\\
AMOLF\\
Amsterdam, Netherlands\\
{\tt\small n.orlov@amolf.nl}
}

\begin{document}
\maketitle
\begin{abstract}
This paper presents a novel approach for denoising Electron Backscatter Diffraction (EBSD) patterns using diffusion models. We propose a two-stage training process with a UNet-based architecture, incorporating an auxiliary regression head to predict the quality of the experimental pattern and assess the progress of the denoising process. The model uses an adaptive denoising strategy, which integrates quality prediction and feedback-driven iterative denoising process control. This adaptive feedback loop allows the model to adjust its schedule, providing fine control over the denoising process. Furthermore, our model can identify samples where no meaningful signal is present, thereby reducing the risk of hallucinations. We demonstrate the DIFFRACT - the successful application of diffusion models to EBSD pattern denoising using a custom-collected dataset of EBSD patterns, their corresponding Master Patterns, and quality values.
\end{abstract}    
\section{Introduction}
\label{sec:intro}

EBSD (Electron Backscatter Diffraction) is an electron microscopy technique for studying crystallographic properties at the nanoscale. In this method, electrons are diffracted by crystallographic planes, producing patterns that reveal crystal orientation. Its high resolution provides detailed information on grains, even at the nanometer scale, offering insights into processes such as degradation mechanisms, electronic properties, and other material characteristics.

\begin{figure}
    \centering
    \includegraphics[width=0.8\linewidth]{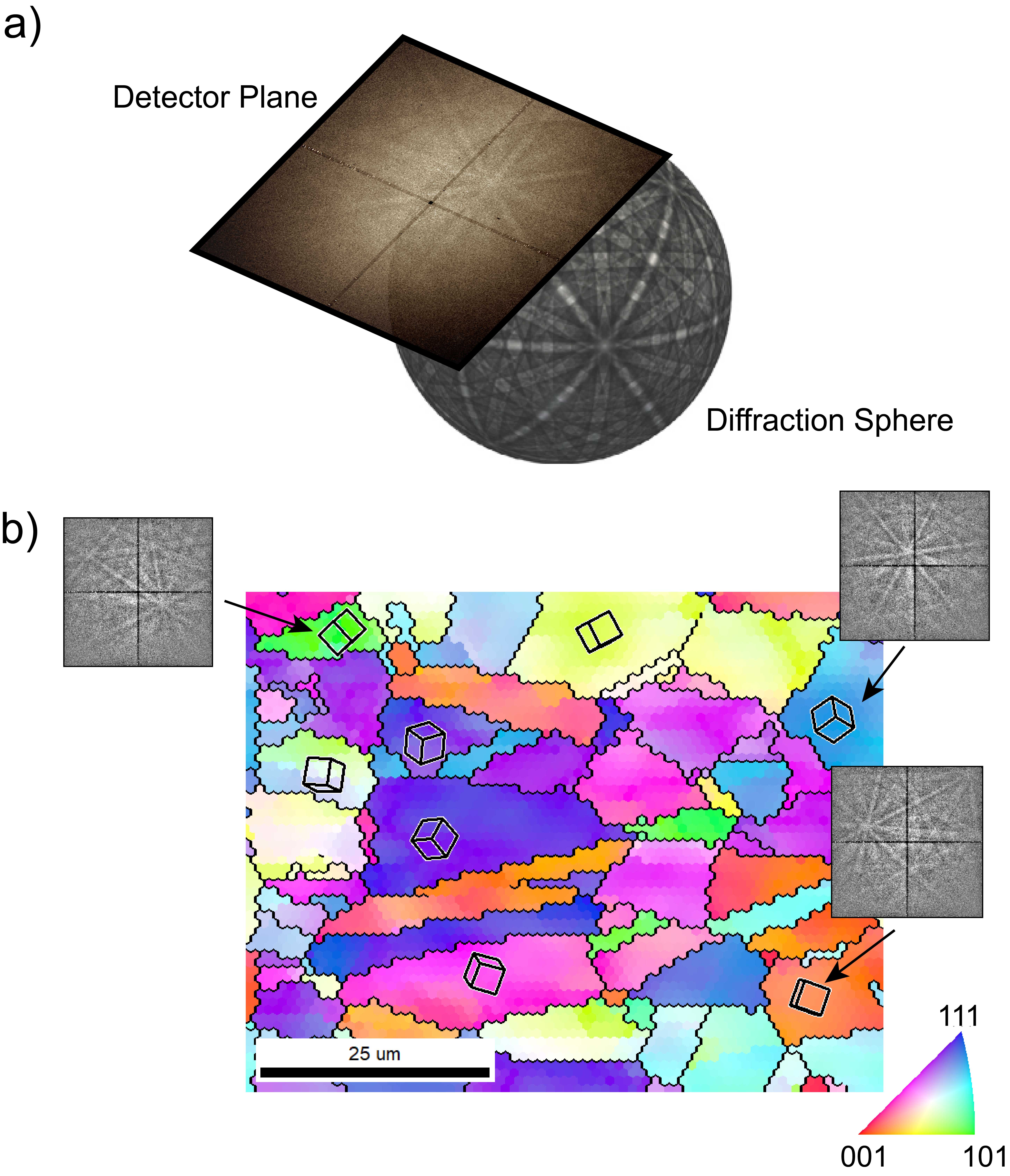}
    \caption{a) Pattern formation as a tangential projection of diffraction sphere on a detector plane; b) Inverse Polar Figure (IPF) map visualising the orientation of a grain by the color of the crystal face pointing perpendicular to the image plane. Each point of the IPF map is the result of EBSD pattern indexing.}
    \label{fig:Fig_1}
\end{figure}

EBSD patterns are grayscale images with diffraction lines called Kikuchi lines, each representing a portion of the diffraction sphere. Similar to maritime navigation using constellations, Kikuchi patterns identify the crystallographic orientation of individual crystallites (Figure \ref{fig:Fig_1}). This process, known as indexing, assigns crystallographic indexes to each Kikuchi line in the pattern.

The quality of EBSD results depends on several factors, including microscope settings (probe current, accelerating voltage, working distance, magnification), detector type (e.g., CCD or single-electron), and, most importantly, the sample itself (surface roughness, composition, crystallization degree, chemical and thermal stability) \cite{Chen2012, Sun2020}. Well-crystallized materials like steel or alumina produce clear Kikuchi patterns \cite{Katrakova2002, Katrakova2001}, whereas beam-sensitive materials such as hybrid lead halide perovskites (promising for solar cell active layers) or metals in thin film form often yield weak, noisy patterns \cite{Jariwala2019, Sun2020}.

Analyzing weak patterns is challenging, with three main approaches: Hough Indexing (HI) \cite{Hough1962, Krieger1992}, Dictionary Indexing (DI) \cite{Chen2015}, and the recently developed Spherical Indexing (SI) \cite{Lenthe2019}. HI processes patterns in Hough space but depends on high-quality data, limiting its use for beam-sensitive materials. DI improves precision for noisy patterns by comparing them to simulated Master Patterns (Figure \ref{fig:Fig_2}) but is computationally intensive. SI also uses a Master Pattern but matches via spherical functions rather than dot products, enabling faster analysis and higher precision than DI.

However, SI’s effectiveness depends on preprocessing, which is subjective, variable, and often requires experience and, at times, luck. While SI can index very noisy patterns lacking visible Kikuchi lines, assessing indexing accuracy remains difficult. In this context, EBSD pattern restoration can enhance post-analysis in three key ways:

\begin{itemize}
\item reduce exposure time in situations where prolonged signal collection might damage samples, all while maintaining quality;
\item make indexing more reliable, consistent, and less dependent on the preprocessing of patterns;
\item enhance pattern processing to enable indexing of patterns that were previously unindexable.
\end{itemize}

Computer vision approaches offer a promising route for denoising EBSD patterns and extracting more information. Current methods include Autoencoders, GANs, CNNs, and diffusion-based techniques. While the first two have been applied to EBSD, their effectiveness depends on sufficiently high pattern quality to ensure stable results and avoid hallucinations. Diffusion learning provides an advantage with its step-by-step approach, but as shown in the Related Works section, even advanced methods can hallucinate when processing content-simple images like EBSD patterns.

To address these challenges and improve the reliability of EBSD pattern denoising, this paper aims to develop a control mechanism that limits the restoration process while preserving vital information in EBSD images. Our contributions include:
\begin{itemize}
    \item We introduce a feedback-driven, adaptive denoising approach that integrates quality prediction to control the iterative process. To our knowledge, this is the first successful application of diffusion models to EBSD pattern denoising.
    \item Our model incorporates a mechanism for distinguishing between meaningful signal and noise, helping to flag potential hallucinations.
\end{itemize}
\section{Related Works}
\label{sec:related}

Image denoising with Deep Learning has been actively developed for over a decade, with most methods based on convolutional architectures. NAFNet is notable as the first nonlinear-activation-free network and has been applied even to medical image denoising, reducing hallucination risk \cite{NAFNetmedical, NAFNet, NAFNetMRI}.
Another example is the Context Guided Network for Semantic Segmentation (CGNet) \cite{CGNetoriginal, CascadedGaze}, which captures both local details and global context through a cascaded gaze mechanism.

Although strong baselines for image denoising, these methods have limitations in scientific imaging, especially microscopy, where long-distance pixel relationships are critical. For images such as EBSD patterns, approaches like GANs, transformers, and diffusion networks (discussed later) are often better suited.

Improving EBSD pattern (EBSP) quality with Deep Learning has been an active research topic in recent years. While diffraction image denoising often adapts techniques from real-world images, it differs due to distinct noise characteristics (less Gaussian, more Poisson) and the unique physics of image formation, involving reciprocal space and greater emphasis on fine details.

Since the pattern dataset forms the foundation for the entire indexing process, most studies first restore EBSD patterns with DL methods, then index the restored patterns and generate IPF maps.

Early EBSD denoising studies used UNet-based autoencoders \cite{Andrews2023}, achieving good restoration quality. At the time, SI was not widely adopted, and many initial patterns were poorly indexed using HI. Today, SI can easily index patterns of that quality, so much of this earlier work focused on improving strain analysis, which demands higher-quality patterns.

Few studies address pattern restoration when images are mostly noise and nearly devoid of visible lines. A notable exception is \cite{Krishna2023}, where patterns are almost lost but experienced users can still spot some Kikuchi lines. The scarcity of such research stems from the difficulty of verifying prediction accuracy and the challenges of collecting ground truth data.

The key advantage of diffusion models for image restoration is their stepwise handling of noisy images. Unlike classical models that map a noisy sample directly to a clean one, diffusion models progressively reduce noise over several steps, particularly valuable for scientific images shaped by physical sample properties. This gradual restoration occurs in image space, transitioning from noisy to clean, and was originally proposed as a general method for image restoration without training separate models for each task \cite{Chang2017}.

The diffusion models are represented by several types of approaches. The Denoising Diffusion Probabilistic Model introduced by \cite{Ho2020} uses Markov chains to model the forward and reverse process. Step by step it applies noise during forward process. Then in backward process the model is learning noise distribution parameters such as mean $\mu_{\theta}(x_t, t)$ and variance $\beta_{\theta}(x_t, t)$ to restore image from noisy state back to original state. 

Despite their general popularity and proved efficiency in denoising real-world imagery data, like photographs \cite{Real-world-enoising}, the diffusion models are still not widely used for scientific image and especially EBSD images denoising. The crucial difference in the \cite{Real-world-enoising} work is in method of sampling, which has helped stabilize the image restoration process. In previous works, every next image in the denoising pipeline is a combination of restored image from the previous state and normalized noise which is left after subtraction of restored image from the noisy one. In the \cite{Real-world-enoising} work, every next image is an image from the previous step and subtracted normalized difference between original noisy image and a restored image from the previous step.

As mentioned above, the main challenge for all denoising approaches is ensuring the stability of the image restoration process. In \cite{Chung2024} was demonstrated, that the restored image is not a single fixed image but rather a variety of potential images, forming a data manifold. A promising approach suggested by \cite{Enhancing-Sample} involves using an adaptive level of noise during each step of the reverse process, rather than applying a constant noise level. This adaptive noise level is estimated at each step based on the distance of the restored image $x_t$ from the noiseless image $x_0$. To facilitate this, a separate network has been introduced to learn the level of noise correction, which is defined as follows: $ \hat{\sigma}_t := \sigma_t[1 + \hat{r}_t] \approx \text{dist}_\kappa (x_t) / \sqrt{n} $.

Another approach to improving EBSD results, presented in the literature, involves processing EBSP using traditional imaging techniques, followed by enhancing the resulting Inverse Pole Figure (IPF) maps through Deep Learning. For example, in \cite{Buzzy2024}, IPF maps were restored using a diffusion-based approach. To stabilize the diffusion model, the authors employed Reduced-Order Generalized Spherical Harmonics (ROGSH), which incorporate additional information to improve restoration performance. In contrast, the present work focuses on restoring EBSD patterns directly, as this remains the most precise and reliable method for enhancing EBSD analysis.
\section{Dataset}

There are EBSD datasets available on the Internet (https://zenodo.org/communities/ebsd). However, these datasets were not used in this work as our approach relies on Master Patterns as a ground truth and employs a patterns quality metric, which is not available in the public data. Therefore, we collected new datasets and generated corresponding pieces of Master Patterns using Spherical Indexing.

We utilized a scanning electron microscope ThermoFisher Verios 460, equipped with an EDAX Clarity detector utilizing a 4-chip Timepix detector offering noiseless digital integration and high detection efficiency \cite{Jakubek2007} from Amsterdam Scientific Instruments (ASI). The data patterns were gathered using EDAX APEX software with beam currents ranging from 50 to 200 pA and accelerating voltages 5-20 kV. Pixel integration times varied from 30 to 250 ms, with different binning options applied (1, 2, 4, and 5 pixels). We analyzed multiple samples (Nickel ($Ni$), Ruthenium ($Ru$), Gold ($Au$), Silver ($Ag$)) to apply a diverse set of measurement conditions and to include various crystallographic variables, including different crystal lattice parameters and crystal systems (Cubic, Hexagonal). In total, the number of experimental patterns used for training was over 163000, with further 16500 patterns used for validation and over 6000 used as a hold-out test set.

\subsection{Master Patterns}

Master Patterns for each crystal structure were generated using EDAX OIM 9 software. This software was also utilized to process and index experimental and restored EBSD patterns using Spherical Indexing. The Master Patterns were used as ground truth for the models training and evaluation.

\section{Methods}

The data in this task differs from traditional image denoising because it includes both ideal images (Master Patterns) and experimental images with an associated 'confidence index' that quantifies the reliability of the experimental data. Additionally, within each grain, the computed Master Patterns are nearly identical, with only minor variations due to defects or crystal lattice bending. This allows experimental patterns within a grain to be considered noisy samples from the same distribution.

\subsection{Proposed approach}

Our approach combines a diffusion-based denoising model with an integrated quality assessment feedback mechanism. This dual-objective architecture allows the model to simultaneously denoise EBSD patterns while evaluating the denoisability and quality of the input patterns. Critically, our model includes a mechanism to identify and flag patterns that represent pure noise rather than noisy observations of actual crystal structures, thus preventing hallucination artifacts.

\begin{figure}  
    \centering  
    \begin{subfigure}{0.9\linewidth}  
        \centering
        \includegraphics[width=1.0\linewidth]{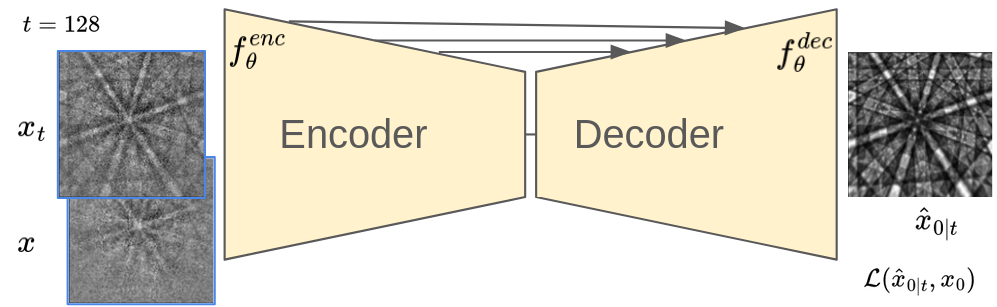}  
        \caption{Stage 1: Training the denoising neural network. The example illustrates the process for $t=128$.}  
        \label{fig:alg1}  
    \end{subfigure}  
    \\ 
    \begin{subfigure}{0.9\linewidth}  
        \centering
        \includegraphics[width=1.0\linewidth]{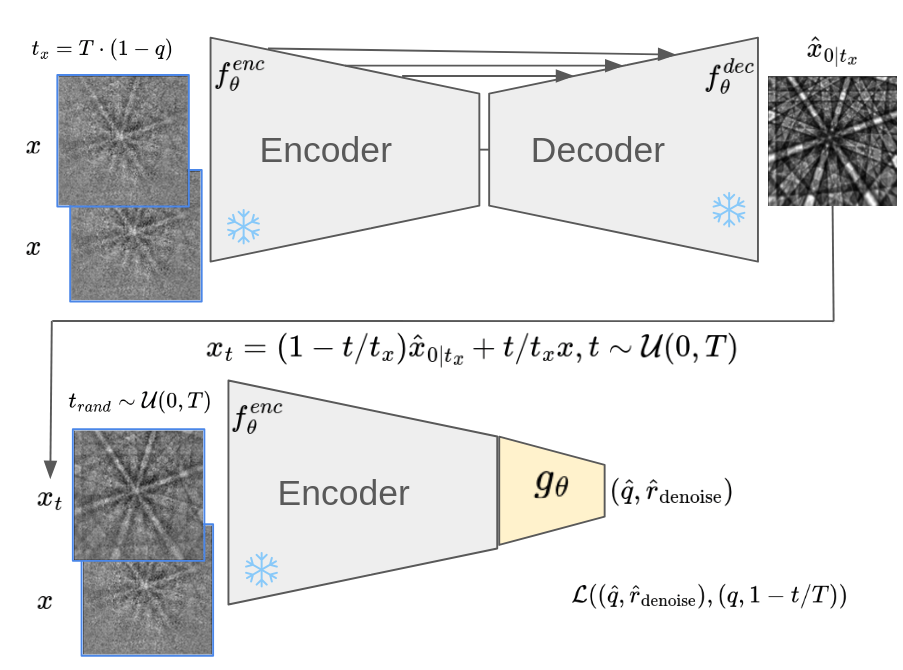}  
        \caption{Stage 2: Training the auxiliary head to predict both the original image quality and denoising progress $r_{denoise}$. Instead of applying an external noise scheduler (as in Stage 1), the pretrained denoising network is used to provide partially denoised images. The function $g_\theta$ is trained using a random value $t_{rand}$ to decouple denoising stage prediction from step number embedding.}  
        \label{fig:alg2}  
    \end{subfigure}
    \caption{Proposed two-stage training procedure for image denoising and quality assessment.}  
    \label{fig:training}  
\end{figure}

\subsection{Training procedure}

Model training was performed in two stages.

\textbf{Denoising model}. On the first stage we training the main model to perform image denoising. The pseudocode for this training procedure is presented in Algorithm \ref{alg:training1}, which is depicted in Figure \ref{fig:alg1}. During training, the network receives a noisy observation \(x_t\) as input and attempts to predict the corresponding clean image \(x_0\). The denoising process is guided by a quality score \(q\) and a noise scale \(s_{\text{noise}}\), with additional Gaussian noise injected at each diffusion step to simulate varying levels of corruption. The noise injection process follows a diffusion model, where the noisy observation \(x_t\) is computed as a weighted combination of the clean image \(x_0\), the noisy input \(x\), and an additional noise term \(\omega_t\) drawn from a standard Gaussian distribution \(\mathcal{N}(0, I)\), Eq. \ref{eq:x_t}.

\begin{equation}
  x_t = (1 - \alpha) x_0 + \alpha x + \beta \omega_t
  \label{eq:x_t}
\end{equation}

The weight \(\alpha\) is determined by the current diffusion step \(t\), relative to the total number of diffusion steps \(T\), as \(\alpha = \frac{t}{t_x}\), where \(t_x = T \cdot (1 - q)\). The noise scale \(\beta\) is determined by the factor \(m \cdot \frac{s_{\text{noise}} \cdot t}{T}\), where \(m\) is drawn from a Bernoulli distribution with probability \(p_{\text{noise}}\), indicating whether additional noise should be applied, while \(s_{\text{noise}}\) controls the level of additional image degradation.

Notably, the algorithm allows the generation of images that can be more degraded than the observed image \(x\), since for each experimental image the current step \(t\) can exceed \(t_x\), leading to simulated further corruption of the image. This, together with the extra noise enables the model to recover clean images from increasingly corrupted data.

On each step, the network has access to the interpolated noisy image of the step \(x_t\) and is guided by the original experimental noisy image \(x\), which gives the network more context (both are stacked as a two channel image), and the current step \(t\).

\begin{algorithm}
\caption{Denoising Model Training}
\label{alg:training1}
\begin{algorithmic}[1]
\Require {Encoder-decoder neural network $f_\theta$, dataset $\mathcal{D} = \{(x_0^{(i)}, x^{(i)}, q_i)\}_{i=1}^N$, where $x_0^{(i)}$ is the clean image (Master Pattern), $x^{(i)}$ is the noisy observation, and $q_i$ is the quality of $x^{(i)}$; total number of diffusion steps $T$, noise probability $p_{\text{noise}}$, and noise scale $s_{\text{noise}}$.}
\While{not converged}
    \State Sample $(x_0, x, q)$ from $\mathcal{D}$
    \State Calculate diffusion step of $x$: $t_x = T \cdot (1 - q)$
    \State Compute interpolation $x_t = (1 - \alpha) x_0 + \alpha x$, where $t \sim \mathcal{U}(0, T)$, $\alpha = \frac{t}{t_x}$,
    \State Generate additional noise: $\omega_t \sim \mathcal{N}(0, I)$
    \State Add noise: $x_t = x_t + \beta \omega_t$
    where   $\beta = m \cdot \frac{s_{\text{noise}} \cdot t}{T}, \quad m \sim \text{Bern}(p_{noise})$.
    \State Normalize: $x_t \leftarrow \frac{x_t - \mathbb{E}[x_t]}{\text{std}(x_t)}$
    \State Predict the clean image: $\hat{x}_{0|t} \leftarrow f_\theta(x_t, x, t)$
    \State Update parameters of $f_\theta$ to minimize loss: $\mathcal{L}(\hat{x}_{0|t}, x_0)$
\EndWhile
\end{algorithmic}
\end{algorithm}

\textbf{Quality Values Regression}. In the second stage of training (Algorithm \ref{alg:training2}, depicted in Figure \ref{fig:alg2}), we perform quality regression using the trained denoising network in a frozen state. Specifically, the bottleneck feature map, which captures an internal representation of the input tensor, is passed through an auxiliary head for the two-values regression task. The goal is to predict two values: the original image quality \(\hat{q}\), and the progress of the denoising process \(\hat{r}_{\text{denoise}}\).

The first predicted value, \(\hat{q}\), represents the estimated quality of the original image at the current denoising step, corresponding to the same ground-truth value \(q\) from the dataset for all diffusion steps. Since the experimental noisy image \(x\) is part of the input tensor at each step, the information necessary for predicting this objective value is present in the input. \(q = 0\) for pure noise images.

The second predicted value, \(\hat{r}_{\text{denoise}}\), estimates the progress of the denoising process. This value depends on the current diffusion step \(t\) and provides a measure of how far along the diffusion process is in recovering the clean image. More specifically, \(\hat{r}_{\text{denoise}}\) estimates how much noise has been removed by the current denoising step. The target value for \(\hat{r}_{\text{denoise}}\) is constructed using the following rule:

\begin{equation}
    r_{\text{denoise}} = \begin{cases} 
    1 - \frac{t}{T}, & \text{if } x \text{ is not noise}, \\
    0, & \text{otherwise.}
    \end{cases}
  \label{eq:r}
\end{equation}

When the image is not fully corrupted (i.e., \(x\) is not pure noise), the value is inversely proportional to the diffusion step, with lower \(t\) values indicating a more advanced stage of denoising; \(r_{\text{denoise}} = 0.0\) for pure noisy images. The sample \(x\) can represent pure noise in two cases: the dataset includes intentionally collected pure noise EBSD patterns, where noise collected as instrumental noise without a sample material installed, or synthetic noise samples generated during training by augmenting the data with pure noise drawn from a standard Gaussian distribution \(\mathcal{N}(0, I)\).

The training of the auxiliary head involves direct estimation of the partially denoised image at diffusion step \(t\):

\begin{equation}
    x_t = (1 - \frac{t}{t_x}) f_\theta(x, x, t_x) + \frac{t}{t_x} x.
\label{eq:x_t_quality}
\end{equation}

This component differs from the procedure in Algorithm \ref{alg:training1}, as the intermediate image state is generated using the actual denoising model rather than a linear interpolation scheduler, utilising the Master Patterns. Since the encoder's input includes the diffusion step \(t\), and since \(\hat{r}_{\text{denoise}}\) is a function of \(t\), it is crucial to decouple the image generation process during training from the estimation of \(\hat{r}_{\text{denoise}}\). This is done as follows: first, \(t \sim \mathcal{U}(0, T)\) is used to construct \(x_t\) and the target value for \(\hat{r}_{\text{denoise}}\). Then during the forward pass, which updates the weights of \(g_\theta\), a different random step \(t_{\text{rand}} \sim \mathcal{U}(0, T)\) is used as input to ensure that the auxiliary head relies on the actual visual features rather than the information from \(t\)-embedding, allowing the value to be used for denoising progress assessment.

As will be shown in the following sections, this two-values regression enables the model to estimate both the initial step at which the denoising process should begin and provide a feedback on the progress of the denoising process.

\begin{algorithm}
\caption{Quality Regression Training}
\label{alg:training2}
\begin{algorithmic}[1]
\Require {Frozen encoder($f_\theta^{enc}$)-decoder($f_\theta^{dec}$) denoising neural network $f_\theta$, auxiliary quality prediction head $g_\theta$, dataset $\mathcal{D}$, total number of diffusion steps $T$.}
\While{not converged}
    \State Sample $(x, q)$ from $\mathcal{D}$
    \State Calculate diffusion step of $x$: $t_x = T \cdot (1 - q)$
    \State Get partially denoised image: $x_t = (1 - \alpha) f_\theta(x, x, t_x) + \alpha x$, where $\alpha = \frac{t}{t_x},  t \sim \mathcal{U}(0, T)$
    \State Set the target $r_{\text{denoise}}$ as per Eq. \ref{eq:r}, using the same $t$ as used for $x_t$.
    \State Predict regression values: $(\hat{q}, \hat{r}_{\text{denoise}}) \leftarrow g_\theta(f_\theta^{enc}(x_t, x, t_{\text{rand}}))$, where $t_{\text{rand}} \sim \mathcal{U}(0, T)$.
    \State Update parameters $g_\theta$ to minimize loss: $\mathcal{L}((\hat{q}, \hat{r}_{\text{denoise}}), (q, r_{\text{denoise}}))$
\EndWhile
\end{algorithmic}
\end{algorithm}

\subsection{Prediction}

\begin{figure}[!t]
    \centering  
    \includegraphics[width=\columnwidth]{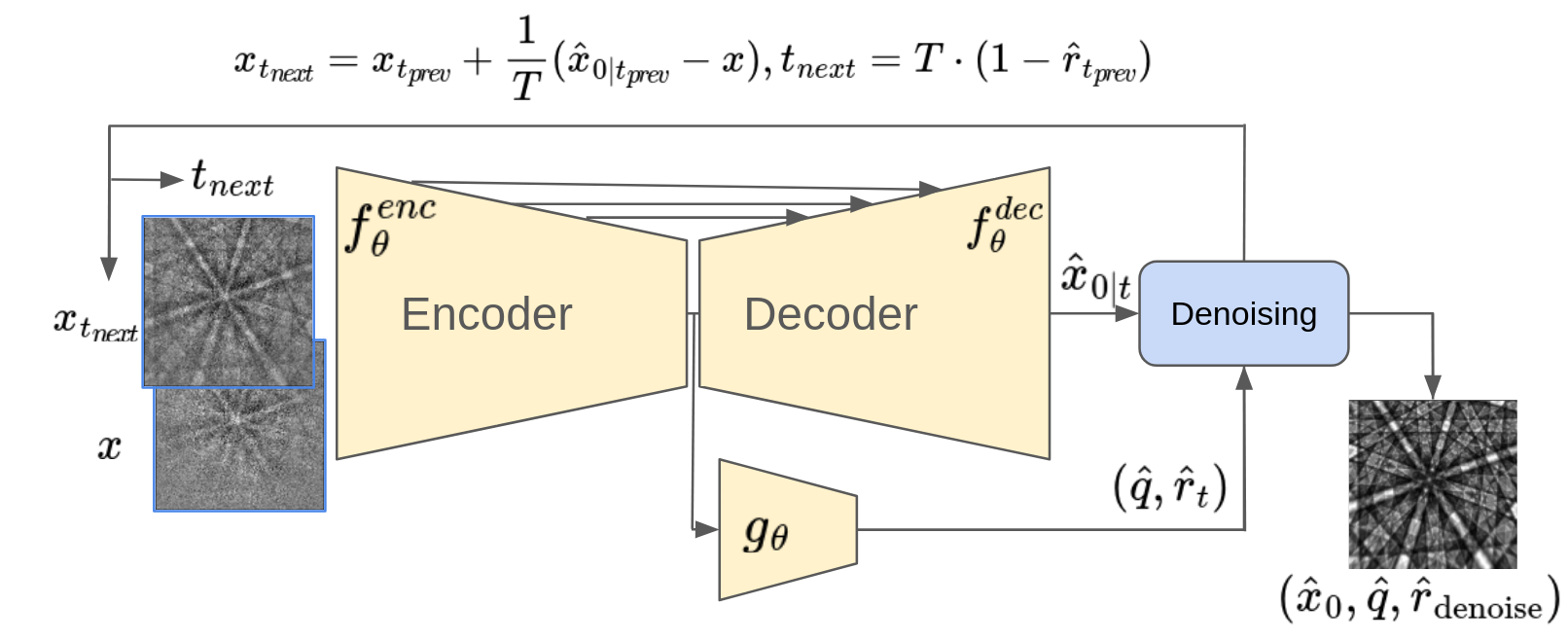}  
    \caption{Diagram of the proposed denoising algorithm with feedback.}  
    \label{fig:alg3}   
\end{figure}

Conceptually, the prediction Algorithm \ref{alg:prediction}, depicted in Figure \ref{fig:alg3}, is similar to the improved sampling algorithm from \cite{Real-world-enoising}, with two key novel features arising from the use of the two regression values. First, we estimate the initial diffusion step based on a dry-run of the neural network (the input timestep is set to \( T \) during the run). 

Second, in contrast to dynamic methods like \cite{Enhancing-Sample}, which adjust the amount of noise removed in each step based on a monotonous denoising process with predicted noise level, we adaptively select the next diffusion step during denoising based on the predicted denoising progress \( \hat{r}_{t_x} \). This adaptive approach allows our algorithm to dynamically adjust the denoising steps: it can repeat a step if further denoising is required or skip steps if the progress indicates sufficient improvement. To prevent infinite loops, the total number of denoising steps is capped at \( T \), and the predicted value \( \hat{r}_{t_x} \) provides a stopping criterion for the process, allowing it to exit at a desired denoising level \( R \) when required by the application.

Moreover, given the definition of \( \hat{r}_{t_x} \), its value approaching zero can serve as a useful indicator to flag potential hallucinations or unreliable predictions. This is particularly valuable in ensuring the robustness of the denoising process and detecting situations where the model might be making implausible predictions. Usage of \( \hat{r}_{t_x} \) allows getting the best denoised image, based on the evaluation, rather then conventionally selected image from the last denoising step.

\begin{algorithm}
\caption{Denoising Prediction with Feedback}
\label{alg:prediction}
\begin{algorithmic}[1]
\Require {Neural network $E_\theta$, consists of encoder($f_\theta^{enc}$)-decoder($f_\theta^{dec}$) denoising neural network, auxiliary quality prediction head $g_\theta$, total number of diffusion steps $T$, desired final denoising $R \in [0, 1]$}
\State $(\hat{q}_{\text{init}}, \hat{r}) \leftarrow g_\theta(f_\theta^{enc}(x, x, T))$
\State Start diffusion step $t_x = T \cdot (1 - \hat{q}_{\text{init}})$; $x_t = x$
\State Step counter $n \leftarrow 0$
\While{$n < T$ and $t_x > 0$ and $\hat{r}_{t_x} < R$}
    \State $(\hat{x}_{0|t_x}, \hat{q}_{t_x}, \hat{r}_{t_x}) \leftarrow E_\theta(x_t, x, t_x)$
    \State $\alpha_t = t_x/T$; $\alpha_{t-1}=(t_x-1)/T$
    \State $\widetilde{x}_t=(1-\alpha_t)\hat{x}_{0|t_x} + \alpha_t x$
    \State $\widetilde{x}_{t-1}=(1-\alpha_{t-1})\hat{x}_{0|t_x} + \alpha_{t-1} x$
    \State Denoising: $x_{t} = x_t - \widetilde{x}_t + \widetilde{x}_{t-1} = x_t + \frac{1}{T}(\hat{x}_{0|t_x} - x)$
    \State Next step as predicted feedback: $t_x = T \cdot (1 - \hat{r}_{t_x})$
    \State Increment step counter $n = n + 1$
\EndWhile
\State Final $\hat{x}_{0}$ selected as $x_t$ with the highest $\hat{r}_{t_x}$.
\end{algorithmic}
\end{algorithm}

Note that the proposed prediction schedule has no reliance on clean images (Master Patterns).

\subsection{Architecture and Implementation details}

All experiments used \( 128 \times 128 \) images and a UNet style CascadedGaze architecture \cite{CascadedGaze}, following the Real Image Denoising protocol from \cite{CascadedGaze} with a network width of 60. The encoder contains [2, 2, 4, 6] CascadedGaze blocks, the middle layer has 10 NAF blocks \cite{NAFNet}, and the decoder comprises [2, 2, 2, 2] blocks. The auxiliary head is composed of an InstanceNorm layer, followed by two convolutional blocks that each downsample the input by a factor of 2 and incorporate a skip connection. After global‑average pooling, the resulting features are fed into a 3‑layer MLP that employs batch normalization and a dropout with rate of 0.1. The hidden layer sizes are 256 and 128, and the final layer outputs two regression values, which are passed through a sigmoid activation.
Diffusion‑step values are encoded with 128‑dimensional sine–cosine positional embeddings and then projected through a fully connected layer. The SiLU function is used as the non‑linearity in this network. 

We used the L1 loss function for both tasks, trained with the AdamW optimizer \cite{AdamW} with \( \beta_1 = 0.9, \beta_2 = 0.9 \), starting with a learning rate of \( 10^{-3} \) and decays to \( 10^{-7} \) using a cosine annealing scheduler over 96 epochs.

Two augmentations were applied during the first stage training: MixUp \cite{mixup} with a probability of 0.25 and a maximum added pattern weight of 0.25, and 10\% rate of Gaussian noise-only images paired with zero targets. \(s_{\text{noise}}\) and \(p_{\text{noise}}\) were set to 0.5. In Algorithms \ref{alg:training2} and \ref{alg:prediction} \( x_t \) values were normalized to range [0, 1]. $T=384$ in all experiments.

The model was implemented in PyTorch 2.6 with mixed-precision training (bfloat16), and experiments ran on two Nvidia A10 GPUs with 24GB VRAM.

\section{Results and discussion}

In this section, we are comparing different training and inference strategies and evaluating the impact of our framework’s components on denoising performance.

\begin{figure*}
    \centering
    \includegraphics[width=1.0\linewidth]{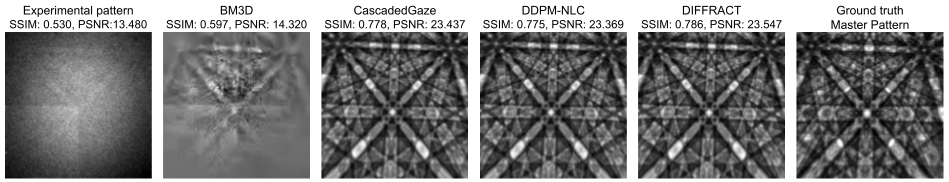}
    \caption{Visual qualitative comparison of different EBSD patterns restoration methods.}
    \label{fig:Fig_7}
\end{figure*}

All experiments use the same fixed data subsets, hyperparameters, and augmentations, with evaluation on a hold-out test set using Peak Signal-to-Noise Ratio (PSNR) and Structural Similarity Index (SSIM).

Since our model can predict a denoised image from any diffusion step, we first compare training strategies for the denoising component. As EBSD denoising has received little attention, we benchmark against representative scientific‑image denoising methods (Table \ref{tab:abl1}), measuring how well a clean image can be reconstructed directly from noisy input. 

\begin{table}
  \centering
  \begin{tabular}{@{}lcc@{}}
    \toprule
    One step process & PSNR ($\uparrow$) & SSIM ($\uparrow$) \\
    \midrule
    Raw data & 17.2198 & 0.5160 \\
    BM3D \cite{BM3D_original, BM3D} & 18.5254  & 0.6310 \\
    NAFNet \cite{NAFNet, NAFNetMRI} & 24.9287 & 0.7991 \\
    CascadedGaze \cite{CascadedGaze} & 24.9309 & 0.7931 \\
    DIFFRACT & 25.0333 & 0.8023 \\
    DIFFRACT with $t_x$ assessment & 25.0337 & 0.8067 \\
    \bottomrule
  \end{tabular}
  \caption{Comparison of different procedures for direct patterns restoration. 
  \textit{Raw data} represents the test dataset without denoising.
  \textit{BM3D}, \textit{NAFNet} and \textit{CascadedGaze} are reference training-free and DL methods.
  \textit{Our method} is similar to the conventional linear algorithm from \cite{Real-world-enoising}, expanded by incorporation of input data quality and linear extrapolation to more degraded images.
  \textit{Our method with $t_x$ assessment} follows one step prediction of $\hat{x}_{0|t}$ by Algorithm~\ref{alg:prediction} (lines 1-5).}
  \label{tab:abl1}
\end{table}

\begin{table}
  \centering
  \begin{tabular}{@{}lcc@{}}
    \toprule
    Diffusion process & PSNR ($\uparrow$) & SSIM ($\uparrow$) \\
    \midrule
    DDPM \cite{Ho2020} & 25.1763 & 0.8043 \\
    DDPM-NLC \cite{Enhancing-Sample} & 25.2081 & 0.8132 \\
    \midrule
    Baseline & 25.0620 & 0.8001 \\
     + start step: $t_x = T \cdot (1 - \hat{q})$ & 25.0905 & 0.8012 \\
     + dynamic step: $t_{\text{next}} = T \cdot (1 - \hat{r})$ & 25.2816 & 0.8055 \\
     + best step selection, based on $\hat{r}_{t_x}$ & 25.2825 & 0.8093 \\
    \bottomrule
  \end{tabular}
\caption{The first two rows report the performance of the reference DDPM and DDPM‑NLC models, both applied with deterministic sampling ($\eta = 0$). \textit{Baseline} represents the standard linear denoising approach similar to \cite{Real-world-enoising}. The remaining rows aprogressively add the components of our proposed denoising scheme with the final values demonstrating results of the complete Algorithm~\ref{alg:prediction}.}
  \label{tab:abl2}
\end{table}

BM3D \cite{BM3D, BM3D_original} which is a widely used collaborative filtering non-local image denoising algorithm, performs far below learning-based methods, indicating that our dataset contains noise and structures beyond classical capabilities, despite BM3D’s success on synthetic noise and other imaging tasks \cite{Block-matchingBM3D, BM3D-IAM}. Since our method utilizes the same base architecture as CascadedGaze \cite{CascadedGaze}, which represents a state-of-the-art evolution of the widely used NAFNet convolutional architecture employed in scientific image denoising \cite{NAFNetMRI, NAFNet}, we trained the models with the original training protocol as a reference baseline. The results show that our training strategy outperforms the baseline approach, achieving an increase in both PSNR and SSIM metrics, confirming the value of quality-defined input generation and linear extrapolation to more degraded states. Predicted image quality for $t_x$ selection benefits even direct denoising without diffusion.

\begin{figure*}
    \centering
    \includegraphics[width=\linewidth]{./figs/Fig_6_IPFmap.jpg}
    \caption{a) IPF maps of the original and restored datasets using DIFFRACT, with patterns shown before any image processing or neighboring-pattern averaging. b) Confidence Index (CI) values distribution for IPF map of the original (blue) and restored (orange) patterns.}
    \label{fig:Fig_10}
\end{figure*}

Next, we investigate different inference-time diffusion-based denoising strategies, analyzing how modifications to the step selection process impact image quality (Table~\ref{tab:abl2}). The results demonstrate that dynamically adjusting the diffusion starting step based on predicted quality improves the performance of iterative denoising as well. The approach of dynamically selecting the next diffusion step based on the assessment of the denoising progress $\hat{r}$ proves beneficial and adds to the metric values. Selecting the final denoised pattern using the highest predicted $\hat{r}$ demonstrates the effectiveness of adaptive denoising progress assessment. This result shows that optimal denoising can occur at intermediate steps, where further iterations may actually degrade quality. Adaptive stopping at the optimal step provides a clear advantage over fixed-step methods, emphasizing the importance of quality-based adaptive criteria for achieving optimal reconstruction. Examples of iterative denoising results are presented in Fig.~\ref{fig:Fig_7} and visual comparison of different methods in Fig.~\ref{fig:Fig_8}.

We also compare our method with other diffusion‑scheduling approaches (Table~\ref{tab:abl2}). For these experiments, we used the same architecture as our model and the same data preparation, including artificial image degradation and augmentations, with the only difference being the scheduling. In particular, we benchmark against restoration with DDPM and DDPM‑NLC (Algorithm 1 from \cite{Enhancing-Sample} with $\eta = 0$). The former is a base diffusion approach, whereas the latter is the closest to ours: its adaptive schedule varies the amount of noise at each step, whereas our approach selects the first and the next step number dynamically. Although the DDPM‑NLC approach achieves slightly better SSIM value, it does not provide a straightforward way to stop denoising at a preferred image‑quality level (the relative pattern quality should be preserved to maintain surface contrast differences like grain boundaries in the results after indexing) or to assess raw‑image quality to reduce the risk of hallucination.

The observed results in Fig.~\ref{fig:fig-6} reveal a clear distinction between actual images and instrumental noise samples, based on the predicted values of $\hat{q}$ and $\hat{r}_{denoise}$. For real EBSD patterns, the final predicted denoising progress values are significantly higher, indicating the model's ability to recognize meaningful patterns. In contrast, for noise data, the predicted values remain consistently low, reflecting the absence of structure in the input data. These findings suggest that the predicted values can serve as a criterion for differentiating between actual images and noise, offering a reliable approach for noise detection and filtering in practical applications.

Fig.~\ref{fig:Fig_10} shows IPF maps of Nickel sample measured with extremely low exposure time - 1 ms. Each point’s color and brightness result from EBSD pattern indexing with SI. In the “original” dataset, point 2 is brighter than point 1 due to a weak Kikuchi line in the EBSD pattern (indicated by white arrows) that increases the OIM9 software’s quality parameter Confidence Index (CI) - confidence in crystal orientation assignment (point 1 does not relieve any visible lines). The same line, restored, is visible in the processed pattern. Original dataset patterns were enhanced using classical image processing (Adaptive Histogram Equalization, background subtraction, and neighboring-pattern averaging), whereas restored patterns were processed only by averaging neighboring patterns. Group of pixels with the same hue represent material grains - their size distribution and crystal orientation are the main information extracted from the EBSD analysis. Notably, the restored dataset preserved the surface-relief texture, reflecting the quality-assessment feedback mechanism that maintains differences between initially higher- and lower-quality patterns. This is consistent with the CI histogram (Fig.~\ref{fig:Fig_10}-b), which shows a wider CI distribution for the restored patterns. Thinner dark lines between grains in the “restored” map indicate better indexing quality and more precise grain-size estimation. Moreover, "restored" map reveal more smaller grains through improved indexing of restored patterns. At the end, our method preserves critical surface features and enhances grain boundary clarity, enabling more accurate grain-size estimation and revealing smaller grains, while requiring less preprocessing than classical approaches.

Our ablation study demonstrates that incorporating image quality assessment into the diffusion denoising process yields consistent improvements over baseline methods. The proposed quality-defined control and noisy pattern extrapolation strategies, combined with adaptive timestep selection, collectively improve denoising metrics. The limited performance of traditional methods like BM3D confirms that EBSD pattern complexity necessitates learning-based approaches, justifying the effectiveness of quality-aware training and adaptive denoising strategies for challenging real-world scenarios in scientific imaging.
\section{Conclusion}

We propose the DIFFRACT - a dynamic method for selecting diffusion steps based on predicted pattern quality, $\hat{q}$, and denoised quality, $\hat{r}_{denoise}$. This adaptive strategy: 1) ensures optimal termination of the denoising process; 2) integrates quality-defined control for generating noisy training samples; 3) extrapolating experimental images to more degraded states. To the best of our knowledge, this is the first time such an approach has been applied to EBSD pattern restoration.
 Our approach significantly improves denoising performance compared to baseline methods. The predicted values of $\hat{q}$ and $\hat{r}_{denoise}$ also enable noise-only data detection and filtering. These contributions enhance the denoising process, improving reconstruction quality and robustness.

This method is proved useful in EBSD measurements, allowing for reduced exposure times while sacrificing some pattern quality, and subsequently restoring the information during post-processing. Furthermore, the suggested approach lessens the reliance on the experimentalist's selection of the image processing sequence and could be useful beyond EBSD context.

\section{Acknowledgments}

We would like to acknowledge the partial support of our computational resources through the Dutch national e-infrastructure with the support of the SURF Cooperative using grant no. EINF-9949.

Collecting EBSD patterns is performed at AMOLF, an NWO funded
institute, and is part of the project "Achieving Semiconductor Stability From The Ground Up" (with project number 19459) which is financed by the Dutch Research Council (NWO), Gatan (EDAX), Amsterdam Scientific Instruments (ASI) and CL Solutions.

\newpage

\clearpage
\newpage

{
    \small
    \bibliographystyle{ieeenat_fullname}
    \bibliography{main}
}

\clearpage

\setcounter{page}{1}

\maketitlesupplementary
\vspace{0pt}
\renewcommand{\thefigure}{S\arabic{figure}}

\begin{figure}[H]
    \centering
    \includegraphics[width=0.5\linewidth]{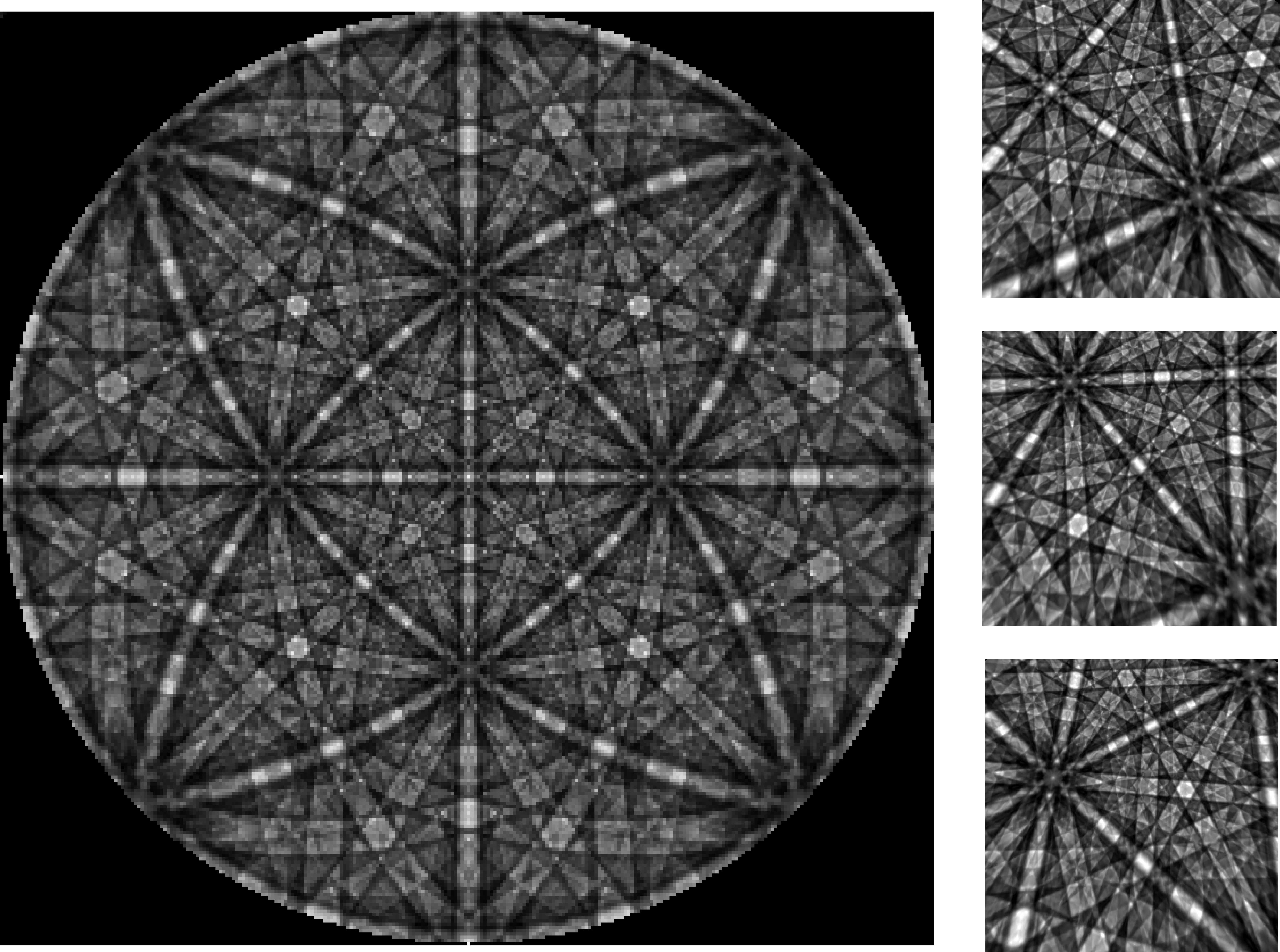}
    \caption{An example of a Master Pattern (MP) and its components identified during Spherical Indexing. The same components of the MP served as ground truth in model training.}
    \label{fig:Fig_2}
\end{figure}

\begin{figure}[H]
    \centering
    \includegraphics[width=0.9\linewidth]{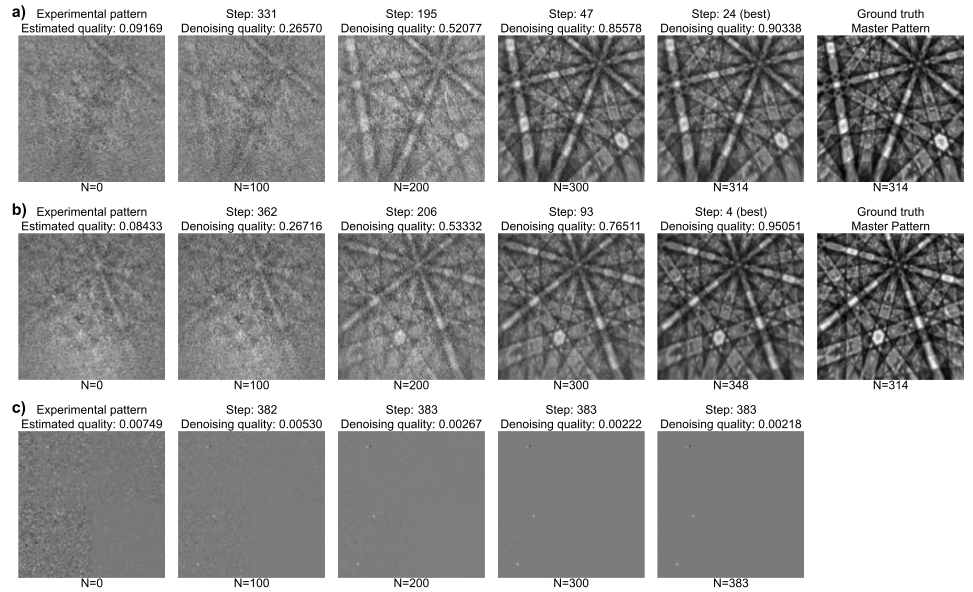}
    \caption{The results of image restoration. $N$ — the number of restoration cycles (maximum $T=384$). Step is the model's diffusion step estimation (lower values indicate a more restored image, according to the model). Initial quality (\textit{estimated quality}) is the predicted quality of the original pattern, $\hat{q}$; \textit{denoising quality} is the model's estimate of the cross-correlation between the restored image and the ground truth, $\hat{r}$. a) and b) - EBSD patterns for Nickel samples of different quality; c) - instrumental noise sample.}
    \label{fig:Fig_8}
\end{figure}

\begin{figure}[H]
    \centering
    \includegraphics[width=0.7\linewidth]{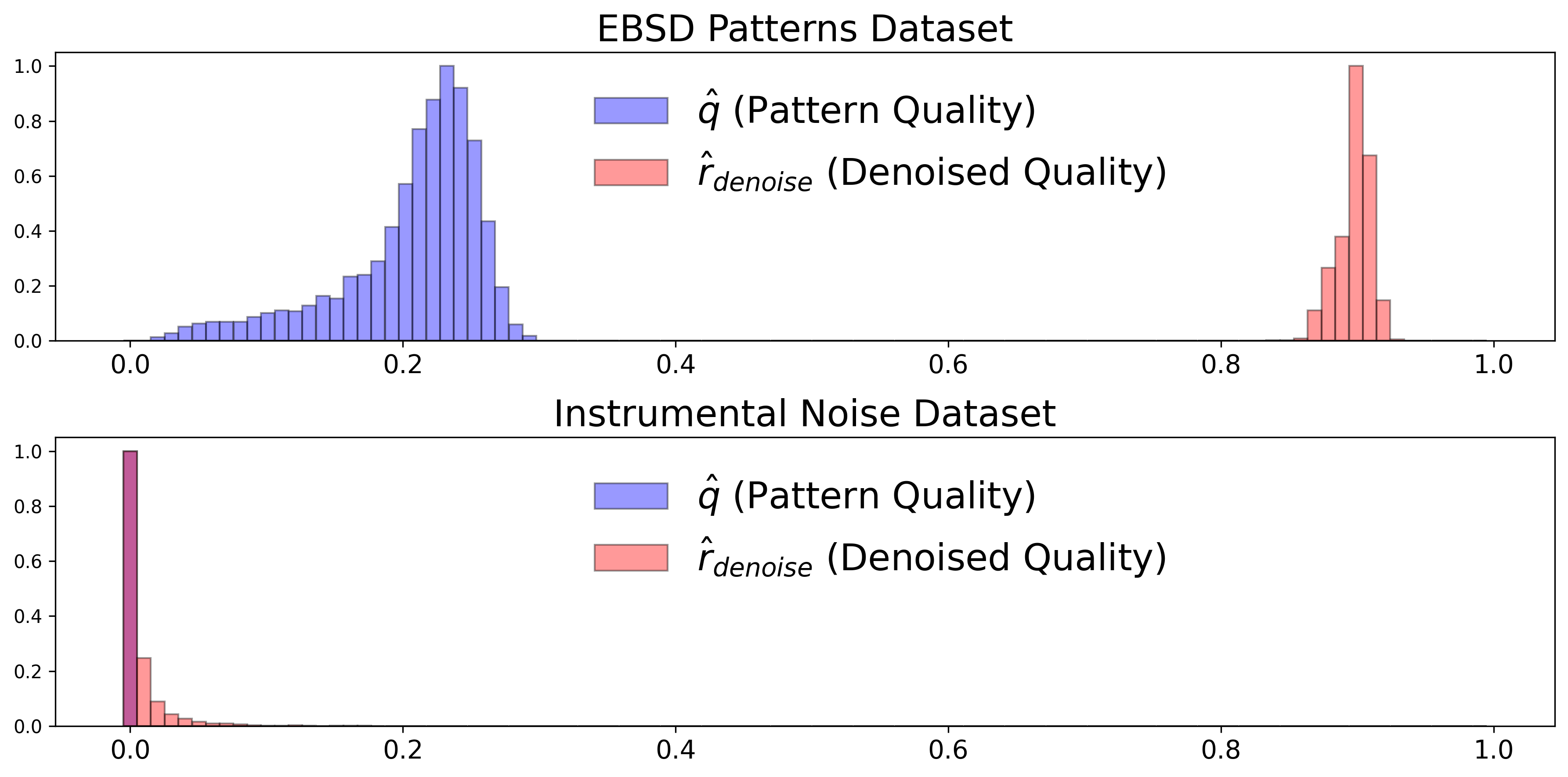}
    \caption{Predicted values of $\hat{q}$ (pattern quality) and $\hat{r}_{\text{denoise}}$ (denoised quality) for a dataset including actual EBSD patterns and images with instrumental noise.}
    \label{fig:fig-6}
\end{figure}

\end{document}